\documentclass[journal = jpcc, manuscript = article]{achemso}
\usepackage{setspace}
\setkeys{acs}{articletitle = true, maxauthors = 100}
\usepackage{graphicx}
\usepackage{amsmath,amssymb,amsfonts,bm}
\usepackage{mathtools}
\usepackage[colorlinks,citecolor=blue,urlcolor=black]{hyperref}
\author{Olugbenga Adeniran}
\affiliation{Department of Chemistry, Wayne State University, Detroit, Michigan 48202, USA}
\author{Sivan Refaely-Abramson}
\affiliation{Department of Materials and Interfaces, Weizmann Institute of Science, Rehovot 7610001,
Israel}
\author{Zhen-Fei Liu}
\email{zfliu@wayne.edu}
\affiliation{Department of Chemistry, Wayne State University, Detroit, Michigan 48202, USA}
\title{Layer-dependent Quasiparticle Electronic Structure  of the P3HT:PCBM Interface from A First-Principles Substrate Screening {$GW$} Approach}
\begin{document}

\begin{abstract}
A prototypical organic photovoltaic material is a heterojunction composed of the blend of regioregular poly(3-hexylthiophene) (P3HT) and [6,6]-phenyl-C$_{61}$-butyric acid methyl ester (PCBM). Microscopic understanding of the energy conversion mechanism in this system involves the relationship between the electronic structure and the atomistic geometry of P3HT:PCBM interfaces. In this work, the effect of the number of P3HT layers on the electronic structure of the P3HT:PCBM interface is studied by means of first-principles $GW$. We apply the substrate screening approach to accelerate such calculations and to better understand the many-body dielectric screening at the interface. The quasiparticle band gap of the entire interface is found to decrease as the number of P3HT layers increases. The gaps of the individual components of the interface are found to be smaller than their isolated counterparts, with strong dependence on the number of P3HT layers. Importantly, when comparing the system of P3HT:PCBM - where a single interface is present - and the system of P3HT:PCBM:P3HT, where an interface is formed on either side of PCBM, we find that the two systems exhibit very different quasiparticle energy level alignments. We discuss possible implications of our findings in related experiments. The observed trends in layer-dependent quasiparticle electronic structure of P3HT:PCBM interfaces provide computational insight into energy conversion pathways in these materials.
\end{abstract}

\section{Introduction}

Organic photovoltaic materials have received tremendous attention over the past two decades owing to the fact that they are durable\citep{Kaltenbrunner2012}, flexible\citep{Kaltenbrunner2012}, inexpensive\citep{Dennler2009}, and highly scalable\cite{Soendergaard2012,Krebs2010}. Among these materials, bulk heterostructures formed from the blend of regioregular poly(3-hexylthiophene) (P3HT) and [6,6]-phenyl-C$_{61}$-butyric acid methylester (PCBM) are widely studied \cite{Hains2010,Kobori2013,Fazzi2017,Menichetti2017} due to their excellent stability\cite{Krebs2009} and ease of synthesis\citep{Dennler2009}. In recent years, several experimental and theoretical efforts have been made to improve the energy conversion efficiency in P3HT:PCBM blends\citep{Krebs2010,Gonzalez2015} and understand underlying mechanisms in such heterostructures\cite{Liu2011}. For the latter, an accurate description of the electronic structure at the heterogeneous interface is a pre-requisite\cite{Kobori2013,DMOB16}. However,  unambiguous experimental determination of the electronic structure of the individual interfaces in such disordered systems is challenging\cite{Jakowetz2017}, due to a high degree of structural variation\cite{Yin2011} in the interpenetrating networks formed between the PCBM and P3HT. As a result, the experimentally determined electronic structure is often an ensemble average of different conformers. In particular, the dependence of the electronic structure on the number of P3HT layers is typically averaged in experiments, and a detailed, quantitative study on the effect of substrate thickness and placement is missing. The lack of definitive structure-property relationships hinders detailed studies of excitonic energy conversion mechanisms such as exciton generation\cite{Vandewal2014}, diffusion\cite{Deotare2015}, and dissociation\cite{Bittner2014}, most of which happen near or at the heterogeneous interface\cite{Grancini2011}. Such processes are expected to strongly depend on non-local dielectric screening effects at the interface.  Therefore, there is a strong need for accurate characterizations of the electronic structure and excitations for conformers of well-defined geometry, as well as quantitative descriptions of the effect of substrate thickness and placement on the interface-related dielectric properties in such complex materials.

First-principles calculations can provide insights into structure-property relationships, complimentary to experimental techniques. In such calculations, representative structures and interfaces can be singled out from the complex interpenetrating network in the P3HT:PCBM  blend, such that their electronic structure can be studied for each configuration with well-defined atomistic geometry and fixed thickness of the substrate. Such information provides guidance for microscopic understanding of the actual complex material. Several theoretical approaches have been used to describe the electronic structure of the P3HT:PCBM heterostructure\cite{Mothy2012,Liu2011,To2014,Kanai2007}, many of which are based on density functional theory (DFT)\cite{Hohenberg1964,Kohn1965}. Although common DFT functionals yield good estimate of the binding energies and geometries\cite{2001}, they typically underestimate the quasiparticle gaps of molecules and semiconductors\cite{Kuemmel2008,Perdew1983,Yang2012} as well as the energy level alignments at heterogeneous interfaces\cite{Neaton2006}. A reliable treatment of the quasiparticle excitations is using the $GW$ formalism\cite{Hedin1965,Hybertsen1986} within the many-body perturbation theory (MBPT). In Ref. \citenum{Li2014}, the band structure and optical properties of the P3HT:PCBM:P3HT junction were studied using $GW$ and the Bethe-Salpeter equation (BSE), particularly focusing on their dependence on the PCBM orientation at the interface. Although charge-transfer excitations in donor-acceptor pairs involving fullerenes were previously studied using the $GW$-BSE approach\cite{Baumeier2012,Duchemin2013}, to the best of our knowledge, Ref.\citenum{Li2014} is the only $GW$-BSE study of the P3HT:PCBM system in the literature. However, a detailed understanding of the many-body screening effect and gap renormalization at the interface is lacking, and the interfacial electronic structure dependence on the thickness or placement of the P3HT layers around the PCBM molecule is still an open question, impeding a comprehensive connection between the quasiparticle  properties and the atomistic structure at the interface. 

In this work, we focus on the many-body effects of the heterogeneous interface by quantitatively describing the gap renormalization of both the adsorbate (PCBM) and the substrate (P3HT). For the adsorbate, such gap renormalization has been well-understood for the case of molecule-metal interfaces in terms of the ``image-charge'' effect\cite{Neaton2006,Thygesen2009,Inkson1971,Lang1973}. Here, we show explicitly that at a molecule-semiconductor interface such as the P3HT:PCBM, the adsorbate and the substrate can mutually screen each other, giving rise to gap renormalization on both sides of the interface compared to the isolated components. To achieve this goal, we perform first-principles $GW$ calculations on the geometry that is found to be energetically most stable (flat-lying\cite{Li2014}) for PCBM on few-layer P3HT. We explore the effect of varying P3HT layers, and compare the system of P3HT:PCBM and the system of P3HT:PCBM:P3HT, to further unveil how the many-body screening of PCBM is drastically affected by the different placements of P3HT layers. We focus specifically on how the valence band maximum (VBM) and the conduction band minimum (CBM) of P3HT are aligned at the interface with the highest occupied molecular orbital (HOMO) and lowest unoccupied molecular orbital (LUMO) of the PCBM, i.e., the so-called level alignment. We apply the substrate screening $GW$  approach recently developed in Ref.\citenum{Liu2019}, which allows the $GW$ calculation of such complex interfaces at a greatly reduced computational cost. This approach assumes that the non-interacting Kohn-Sham (KS) polarizability of the interface can be approximated as the sum of the polarizabilities of the isolated substrate and that of the isolated adsorbate\cite{Liu2019,Ugeda2014,Xuan2019}. Physically, this assumption is valid for weakly coupled systems with negligible orbital hybridization such as the P3HT:PCBM interfaces studied here. This assumption is numerically verified in this work via a comparison to direct $GW$ calculation of the one-layer P3HT:PCBM interface.

\section{Computational Methods}
\label{sec:comp}
\subsection{Modeling and Geometry Optimizations}

\begin{figure}[b!]
\centering
\includegraphics[width=6.5in]{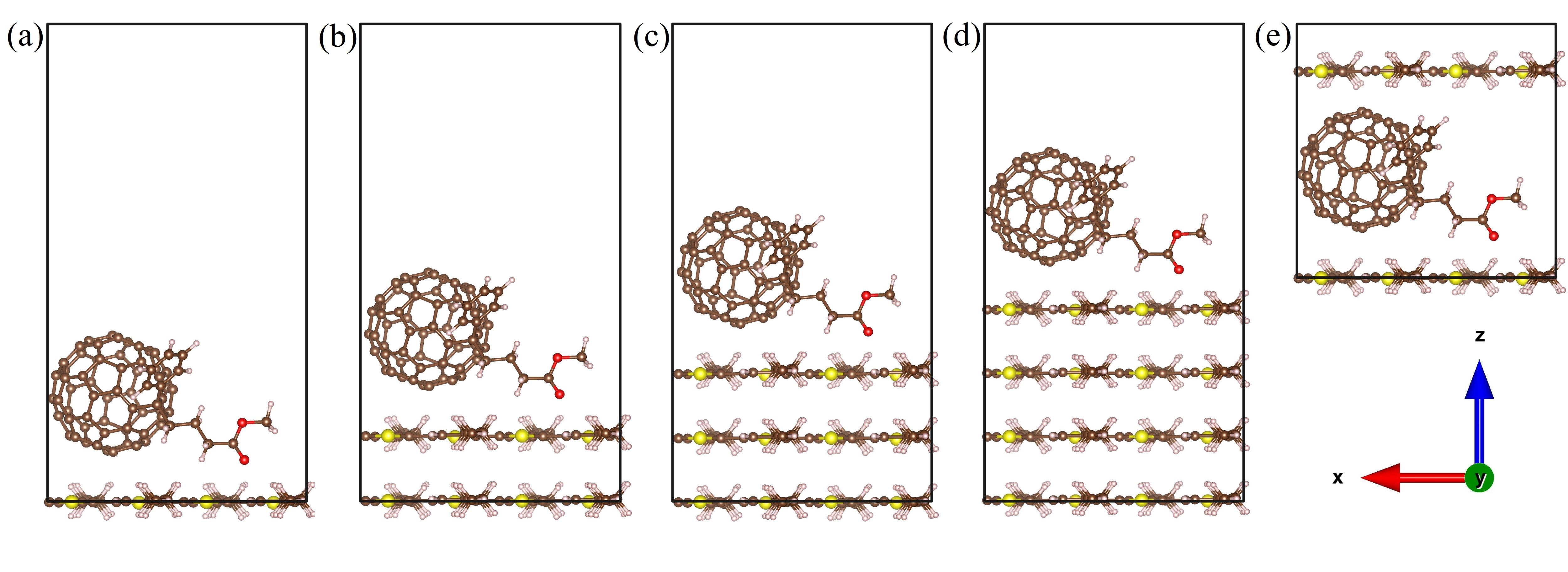}
\caption{Optimized structures of the interface systems. (a) (P3HT)$_1$:PCBM, (b) (P3HT)$_2$:PCBM, (c) (P3HT)$_3$:PCBM, (d) (P3HT)$_4$:PCBM, and (e) (P3HT)$_1$:PCBM:(P3HT)$_1$. The thick borders represent periodic boundary conditions. Interface simulation cells of (a) - (d) have a dimension of 15.94 \AA$\times$15.0 \AA$\times$30.0 \AA, while the interface simulation cell of (e) has a dimension of 15.94 \AA$\times$15.0 \AA$\times$17.17 \AA.  Color code: C - brown, S - yellow, O - red, H - pink. This figure is rendered using VESTA\cite{Momma2008}.}
\label{figure1}
\end{figure}

We first construct interface supercells containing one PCBM molecule adsorbed on one, two, three, and four layers P3HT, respectively. We denote these systems by (P3HT)$_n$:PCBM, where $n$ is the number of substrate layers. Each P3HT layer consists of one linear polymer with four thiophene rings along the backbone (the $x$ direction, see Fig. \ref{figure1}) in the periodic supercell. Our modeling of (P3HT)$_n$:PCBM includes vacuum along the out-of-plane direction, where the supercell size is 15.0 {\AA} along the $y$ and 30.0 {\AA} along the $z$ direction (the surface normal). This setting provides one clear interface between P3HT and PCBM. As a comparison, we also construct an interface supercell containing one P3HT layer on each side of the PCBM molecule [see Fig. \ref{figure1}(e)], which we denote by (P3HT)$_1$:PCBM:(P3HT)$_1$. This setting provides two interfaces formed between P3HT and PCBM and therefore there is no need to use vacuum, similar to the system studied in Ref. \citenum{Li2014}.

The geometry optimizations are performed using DFT as implemented in the \texttt{Quantum ESPRESSO} package\citep{Giannozzi2009}. The optimized norm-conserving Vanderbilt (ONCV) pseudopotentials\citep{Schlipf2015,Hamann2013} are used in all calculations. We first optimize the primitive unit cell of the one-layer P3HT polymer containing two thiophene rings along the backbone. We fully relax both the lattice constant along $x$ and the internal coordinates of all atoms using the Perdew-Burke-Ernzerhof (PBE) functional\cite{Perdew1996}, a \textbf{k}-mesh of $6 {\times} 3 {\times} 1$, and an energy cutoff of 90 Ry. The resulting lattice parameter along $x$ is 7.97 {\AA}. Multi-layer P3HT primitive unit cells are modeled by placing additional layer(s) below the top layer. In order to capture the weak interactions between the layers, all internal coordinates in the multi-layer P3HT unit cells are fully relaxed using the vdw-DF-cx functional\cite{Berland2014} with an energy cutoff of 60 Ry until the residual forces are below 0.05 eV/{\AA}, while keeping the lattice parameters along the in-plane directions the same as that of one-layer P3HT. The resulting distance between two neighboring P3HT layers is about 4.0 {\AA}. The optimized unitcells are then used to build $2 {\times} 1 {\times} 1$  interface supercells to accommodate one PCBM molecule.  

For the (P3HT)$_n$:PCBM interfaces, we adopt the ``flat-lying'' geometry\cite{Li2014} of the PCBM on the substrate (see Fig. \ref{figure1}), with the coordinates of the PCBM molecule fully relaxed. The substrate coordinates are kept fixed in their relaxed positions, to ensure that the subsequent KS polarizability folding is exact. Again, in order to capture the weak interactions at the P3HT:PCBM interface, the vdw-DF-cx functional\cite{Berland2014} is used together with a \textbf{k}-mesh of $3 {\times} 3 {\times} 1$ and an energy cutoff of 60 Ry until all residual forces are below 0.05 eV/{\AA}. The relaxed adsorption height is about 3.0 {\AA} (measured between the bottom of the fullerene ring and the carbon backbone in P3HT) for all interfaces in this work. The optimized (P3HT)$_n$:PCBM interface structures are shown in Fig. \ref{figure1}(a)-(d). For the (P3HT)$_1$:PCBM:(P3HT)$_1$, we start with the relaxed (P3HT)$_2$:PCBM and decrease the amount of vacuum until the distances between the PCBM and both P3HT layers become about the same (3.0 {\AA}). This interface structure is shown in Fig. \ref{figure1}(e).

\subsection{$GW$ Calculations with Substrate Screening}
Due to the large system size and the computational cost of standard $GW$ calculations, we use the substrate screening method\cite{Liu2019} for (P3HT)$_n$:PCBM, and validate it by comparing the results to a direct $GW$ calculation of the entire (P3HT)$_1$:PCBM interface. All polarizability and self-energy calculations are performed using the \texttt{BerkeleyGW} package\cite{Deslippe2012}, at the $G_0W_0$ level. For (P3HT)$_1$:PCBM:(P3HT)$_1$, we only perform a direct $GW$ calculation due to the relatively small system size.

To use the substrate screening method as developed in Ref.\citenum{Liu2019}, the KS polarizability of the substrate unit cell needs to be calculated and then folded in the reciprocal space to that of the supercell. We use a \textbf{q}-mesh of $6 {\times} 3 {\times} 1$ and a 5 Ry dielectric cutoff and a summation of 4900 bands in the calculation of the KS polarizability of the substrate unit cell. A ``$\mathbf{q}$-shifted-grid'' wavefunction, \texttt{WFNq}, required in the polarizability calculations to evaluate the limit of $\mathbf{q}\to 0$, includes 200 bands for one-, two-, and three-layer P3HT and 300 bands for four-layer P3HT. We then fold this quantity in the reciprocal space to generate the substrate polarizability in the $2 {\times} 1 {\times} 1$ supercell that is equivalent to the interface simulation cell. 

The substrate screening method also requires the calculation of the adsorbate polarizability in a simulation cell that has the same dimensions in $xy$ as the interface, but with a much smaller size along the $z$ direction, referred to as ``small-$z$'' cell in Ref. \citenum{Liu2019}. A length of 15 {\AA} - half the size of the interface simulation cell - is used in this adsorbate ``small-$z$'' cell. We use a \textbf{q}-mesh of $3 {\times} 3 {\times} 1$, a 5 Ry dielectric cutoff, and a summation of 4900 bands to calculate the adsorbate polarizability in the ``small-$z$'' cell. The ``$\mathbf{q}$-shifted-grid'' wavefunction, \texttt{WFNq}, includes 200 bands. We then use the real-space mapping procedure of Ref. \citenum{Liu2019} to obtain the adsorbate polarizability in the simulation cell of the interface. 

Once the substrate and adsorbate polarizabilities are obtained, we add these two quantities for each $\mathbf{q}$ point to approximate the polarizability of the interface. We then calculate the dielectric matrix of the interface followed by standard self-energy calculations. The semiconductor screening\cite{Deslippe2012} and slab Coulomb truncation\cite{IsmailBeigi2006} are used for both dielectric function and self-energy calculations. For the interface, we have checked that a 5 Ry dielectric cutoff and 9800 bands converge the self-energies (which corresponds to 4900 bands in both the substrate primitive unit cell and the adsorbate ``small-$z$'' cell). The self-energy is calculated based on the generalized plasmon pole (GPP)\cite{Hybertsen1986} model of the dielectric function as well as the static remainder\cite{Deslippe2013}.

\section{Results and Discussion}

\subsection{P3HT Substrate}

We first show the $GW$ results for the multi-layer P3HT substrates calculated in their primitive unit cells, focusing on how the VBM-CBM gap is affected by the number of layers. We denote the systems by (P3HT)$_n$, with $n$ the number of layers. The (P3HT)$_n$ is periodic along the $x$ direction, so the bands are dispersive only along $x$, and therefore we only consider the gap along the $x$ direction in the Brillouin zone. In our modeling, these primitive unit cells have the same size along $y$ and $z$ directions as those interface systems shown in Fig. \ref{figure1}(a)-(d). 

\begin{figure}[b!]
\centering
\includegraphics[width=3.5in]{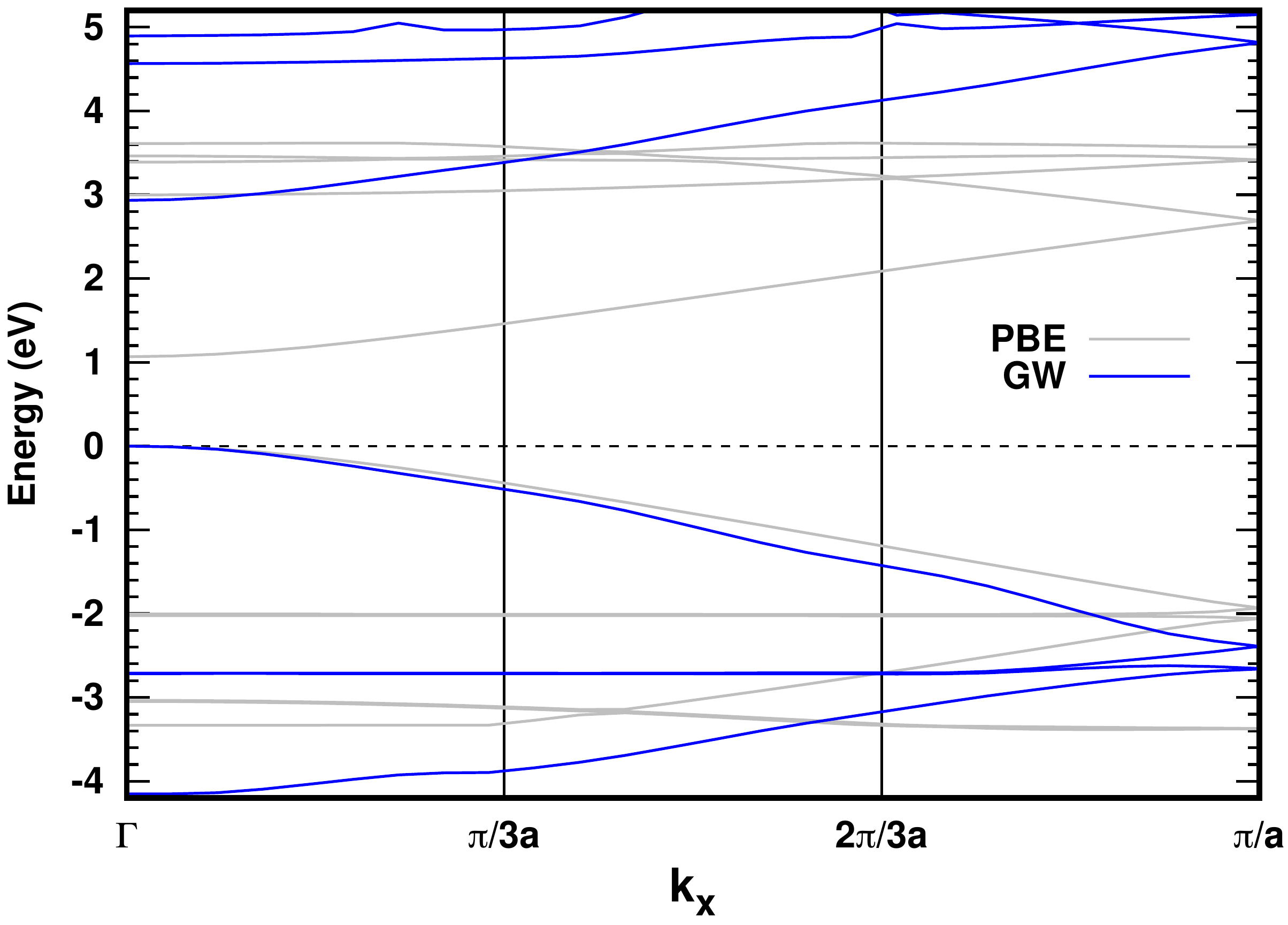}
\caption{Band structure of (P3HT)$_1$, from PBE (gray) and $GW$ (blue). The VBM from both methods are aligned and set to be zero. The bands are dispersive only along the $x$ direction. Only those bands between VBM-4 and CBM+4 are shown.}
\label{fig:bands}
\end{figure}

Fig. \ref{fig:bands} shows the band structure of (P3HT)$_1$, using PBE (gray) and $GW$ (blue), with the VBM from both methods aligned and set to be zero. One can see that $GW$ greatly opens the PBE gap, while preserving the qualitative features of all PBE bands. The top valence band and the bottom conduction band are slightly more ``curved'' in $GW$ than their PBE counterparts, consistent with other materials \cite{GDR19}. Table \ref{tab:table1} shows the gap as a function of $k_x$ for (P3HT)$_n$, from the $\Gamma$ point to the edge of the Brillouin zone $k_x=\pi/a$ with $a$ being the lattice constant along $x$ (7.97 \AA). 

\begin{table}[t!]
\caption{Gaps between the VBM and the CBM at different $\mathbf{k}$ points along the $x$ direction in the reciprocal space, for (P3HT)$_n$ unit cells consisting of different number of layers. $a$ is the lattice constant along the periodic direction. All gaps are in eV and are calculated at the $GW$ level.}
\label{tab:table1}
\begin{tabular}{c|cccc}
\hline
$k_x$      & 0\hspace{0.05in}$(\Gamma)$ & $\pi/3a$ & $2\pi/3a$ & $\pi/a$ \\ \hline
(P3HT)$_1$ & 2.93        & 3.90     & 5.56      & 7.21    \\
(P3HT)$_2$ & 2.28        & 3.22     & 4.83      & 6.48    \\
(P3HT)$_3$ & 1.98        & 2.89     & 4.49      & 6.14    \\
(P3HT)$_4$ & 1.87        & 2.75     & 4.33      & 5.98    \\ \hline
\end{tabular}
\end{table}

The PBE gaps for (P3HT)$_n$ with $n$=1, 2, 3, and 4 are 1.07 eV, 0.75 eV, 0.62 eV, and 0.62 eV, respectively, at the $\Gamma$ point. In Table \ref{tab:table1}, we see that $GW$ significantly opens the PBE gaps, as expected. P3HT is a direct-gap semiconductor regardless of the number of layers, and the band gap is found at the $\Gamma$ point. This is consistent with Ref. \citenum{Oezcelik2020}. We notice that as the number of layers increases, the $GW$ gaps decrease more rapidly than the PBE gaps, due to the many-body screening effect between adjacent layers. This screening is similar in nature to that in a molecular crystal, whose gap is smaller than an isolated molecule\cite{Sato1981,RefaelyAbramson2013}. The $GW$ gap difference between the one-layer and four-layer P3HT is more than 1 eV across the Brillouin zone, suggesting that changing the thickness of this material is an effective way to tune its electronic structure. Additionally, the gap difference between the three-layer and the four-layer P3HT is only about 0.1 - 0.15 eV across the Brillouin zone.  Notably, the gap difference across the series is not prominent from the PBE results without applying $GW$ corrections, due to the missing of non-local correlations in PBE. We can infer from this result that addition of more layers will only marginally further reduce the gap, suggesting that we are approaching the semi-infinite limit for the substrate surface with four P3HT layers. Our calculated band gap for the (P3HT)$_4$ at the $\Gamma$ point (1.87 eV) is in good agreement with other calculations in Ref. \citenum{Li2014} (1.99 eV) and experimental measurements (1.9 eV)\cite{Baran2017,Kroon2008,Hou2006}. These gaps of the isolated multi-layer P3HT will be compared to the P3HT gaps in the interfaces below, and the difference will be attributed to the dielectric screening effect due to the PCBM, a point to be elaborated later.

We note in passing that with the computational convergence parameters used, namely the energy cutoff and the number of empty bands in the dielectric function, the absolute values of the VBM and CBM energies measured with respect to the vacuum may not be fully converged (therefore we do not include them in Table \ref{tab:table1}), but we have checked that the VBM-CBM gaps are converged within 0.05 eV with respect to the dielectric cutoff used in the $GW$ calculations.

\subsection{PCBM Adsorbate}

Based on the substrate screening approach as detailed in Ref. \citenum{Liu2019}, we perform $GW$ calculation of the PCBM in a ``small-$z$'' simulation cell, with the same size along the in-plane directions and half the size along the $z$ direction of the interface simulation cell. For a meaningful comparison between the PCBM adsorbate and the P3HT:PCBM interfaces, we fix the geometry of the molecule in the PCBM calculation as that relaxed in the (P3HT)$_n$:PCBM interfaces. We have checked that the PCBM molecular geometry is largely independent of the number of P3HT layers. The $GW$ gap between the HOMO and the LUMO is calculated to be 4.12 eV, while the PBE gap is 1.51 eV. The bands are essentially dispersionless so we use the term ``molecular orbital'' when mentioning levels associated with PCBM. As a comparison, we also computed the PCBM molecule in the interface simulation cell (i.e. ``large-$z$''), which results in a $GW$ gap of 4.18 eV. Hence, the use of the ``small-$z$'' cell does not significantly change the quality of the $GW$ results, but is central in greatly reducing the computational cost. We emphasize that although these calculations only involve the PCBM molecule, we use periodic boundary conditions along all directions as well as the slab Coulomb truncation \cite{IsmailBeigi2006}, to be consistent with the interface calculations in order to facilitate meaningful comparisons. Physically, this means that we consider a periodic monolayer instead of an isolated molecule.

The computed $GW$ gap for the PCBM molecule quantitatively agrees with the result in Ref.\citenum{Qian2015}, which also employed the $GW$ approach for this molecule. In the literature, the PCBM gap is usually claimed to be around 2 - 2.5 eV \cite{Kroon2008,Li2012,Shih2013}. We point out that this value is likely the gap exhibited in the P3HT:PCBM heterojunction, where the many-body screening significantly reduces its gap compared to the isolated molecule. This value therefore largely depends on the dielectric environment that the PCBM molecule ``feels'', an important point we elaborate on below. We again note in passing that with the convergence parameters used, the absolute values of the HOMO and LUMO measured with respect to vacuum may not be fully converged, but we have checked that the HOMO-LUMO gap is indeed converged within 0.05 eV with respect to the dielectric cutoff used in the $GW$ calculations.

\subsection{P3HT:PCBM Interfaces}

% assignment of orbitals                                                                                                                                                                
To understand how the P3HT levels and PCBM levels are aligned at the heterogeneous (P3HT)$_n$:PCBM interface, we first need to assign interface frontier orbitals as either (P3HT)$_n$ or PCBM orbitals, where $n$=1, 2, 3, and 4. Although one can easily do so by visualizing the shape and localization of interface orbitals, we decide to proceed with a quantitative approach. For each system, we expand isolated (P3HT)$_n$ or PCBM orbitals in the basis of interface orbitals: $\psi_i=\sum_\mu C_{i\mu} \phi_\mu^{\rm interface}$. Here, $\psi_i$ is one of the following: (P3HT)$_n$ VBM, (P3HT)$_n$ CBM, PCBM HOMO, or PCBM LUMO; $\phi_\mu^{\rm interface}$ are the orbitals of the (P3HT)$_n$:PCBM interface. All the orbitals involved in this expansion are PBE KS orbitals. Since the $G_0W_0$ calculations do not change the nature of the orbitals\cite{Hybertsen1986}, the orbital assignments we conclude here carry forward to $GW$ results. When the expansion coefficient $|C_{i\mu}|^2$ is close to unity (indeed the case in this work), the interface orbital $\phi_\mu$ is quantitatively very similar to $\psi_i$, a frontier orbital of the isolated (P3HT)$_n$ or PCBM that we are expanding. Therefore, $\psi_\mu^{\rm interface}$ physically originates from and can be assigned as one of the frontier orbitals of the isolated (P3HT)$_n$ or PCBM. The same analysis applies to (P3HT)$_1$:PCBM:(P3HT)$_1$, where the substrate is taken to be (P3HT)$_2$. 

\begin{table}[t!]
\caption{Identification of P3HT and PCBM frontier orbitals in the (P3HT)$_1$:PCBM interface system, calculated at the $\Gamma$ point. This is achieved via assignment of interface orbitals as either substrate or adsorbate orbitals when the expansion coefficient $|C_{i\mu}|^2$ is close to unity for a given interface orbital $\phi_\mu$. For (P3HT)$_n$:PCBM with $n \geqslant 2$ as well as (P3HT)$_1$:PCBM:(P3HT)$_1$, see Supporting Information.}
\label{tab:table2}
\begin{tabular}{c|c|c}
\hline
                               & interface orbital & $|C_{i\mu}|^2$ \\ \hline
\multicolumn{1}{c|}{P3HT VBM}  & VBM          & 0.989      \\
\multicolumn{1}{c|}{P3HT CBM}  & CBM+3        & 0.902      \\
\multicolumn{1}{c|}{PCBM HOMO} & VBM-1        & 0.995      \\
\multicolumn{1}{c|}{PCBM LUMO} & CBM          & 0.981      \\ \hline
\end{tabular}
\end{table}

Table \ref{tab:table2} shows the assignment of (P3HT)$_1$:PCBM interface orbitals as either P3HT or PCBM orbitals, at the $\Gamma$ point. Results for other $\mathbf{k}$ points are very similar, given the fact that the PCBM is dispersionless. For instance, at $k_x=2\pi/3a$ with $a$ the lattice constant of the interface simulation cell along the periodic direction, the scenario is only different from Table \ref{tab:table2} in that (P3HT)$_1$ CBM corresponds to CBM+4 in the interface. Results for (P3HT)$_n$:PCBM interfaces with $n \geqslant 2$ as well as (P3HT)$_1$:PCBM:(P3HT)$_1$ are shown in the Supporting Information. 

From Table \ref{tab:table2}, we conclude that the (P3HT)$_1$:PCBM interface is a ``type-II''  heterojunction\cite{2010} with staggered band gap: the VBM of the interface is localized on the substrate (P3HT), while the CBM of the interface is localized on the adsorbate (PCBM). Increasing the number of P3HT layers does not qualitatively change this feature (see Supporting Information), which is true at both PBE and $GW$ levels. The ``type-II'' heterojunction is the ground for interfacial exciton dissociation and photovoltaic applications\cite{Lo2011}: both P3HT and PCBM can absorb photons to produce excitons and the staggered band gap at the interface then facilitates the electron (hole) to transfer from LUMO (HOMO) on one side of the interface to the other, forming charge-transfer excitons across the P3HT:PCBM interface\cite{Li2012}. Our predictive $GW$ calculations of the level alignment at the interface therefore offer a quantitative understanding of such charge-transfer mechanisms.

% level alignment for one-layer             

\begin{figure}[b!]
\centering
\includegraphics[width=3.5in]{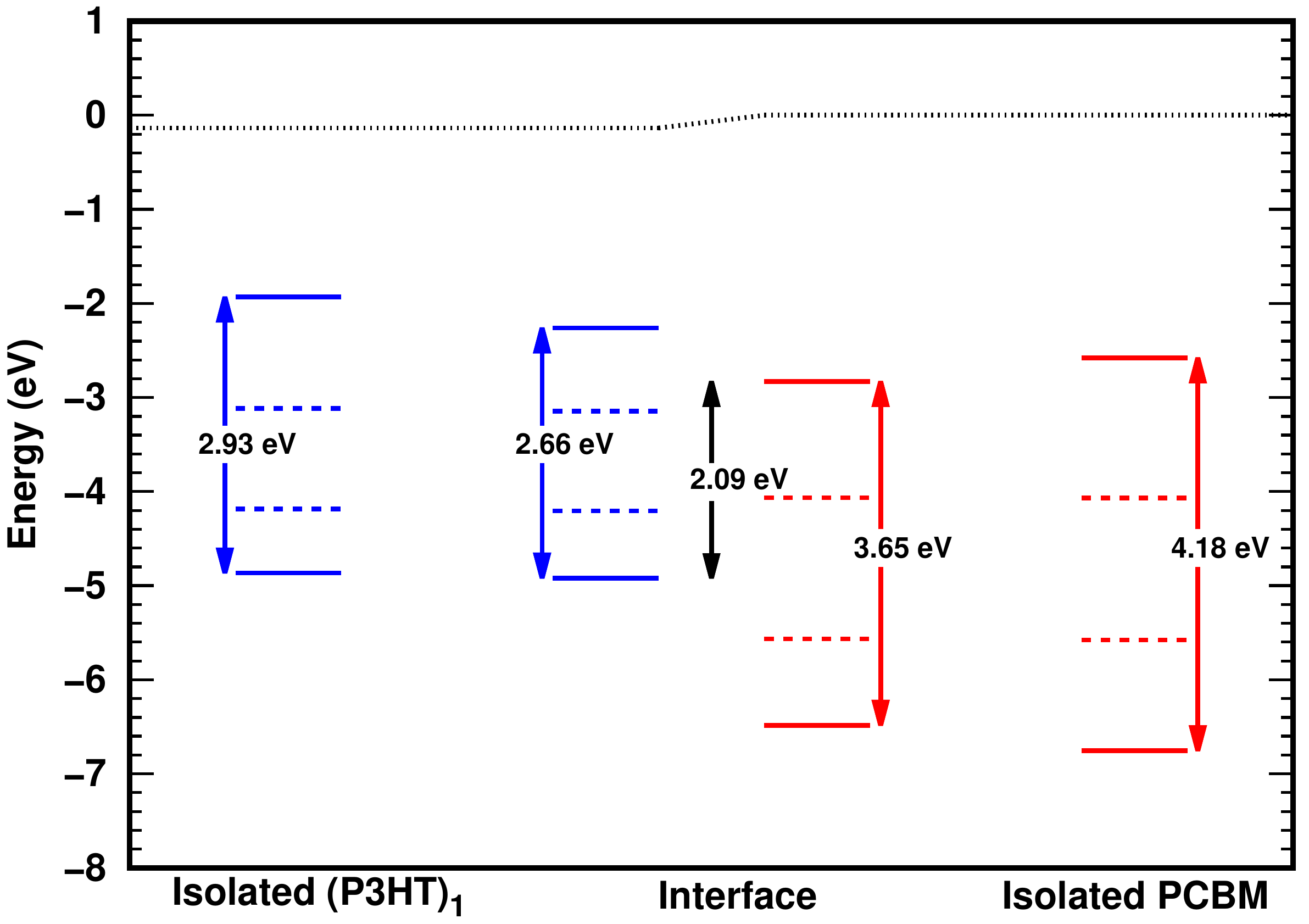}
\caption{Energy level alignment at the $\Gamma$ point for the interface formed between PCBM and one-layer P3HT [as shown in Fig. \ref{figure1}(a)]. Blue (red) lines indicate frontier energy levels of the isolated P3HT (PCBM) or the interface energy levels that are localized on P3HT (PCBM), with the latter discussed in Table \ref{tab:table2}. Blue arrows indicate the VBM-CBM band gaps of P3HT; red arrows indicate the HOMO-LUMO gaps of PCBM; and the black arrow indicates the fundamental gap of the interface system. Solid lines are from $GW$ and dashed lines are from PBE. Vacuum levels on both sides of the interface are shown using dotted black lines. We align the isolated P3HT or PCBM energy levels with their corresponding interface energy levels at the respective vacuum. The vacuum level difference across the interface is a reflection of the interface dipole. All energies are in eV.}
\label{figure2}
\end{figure}

The level alignment diagram (at both PBE and $GW$ computational levels) at the $\Gamma$ point for the (P3HT)$_1$:PCBM interface system is shown in Fig. \ref{figure2}, and the level alignment diagrams for (P3HT)$_n$:PCBM with $n \geqslant 2$ are shown in the Supporting Information. Since the P3HT is a direct-gap semiconductor with the gap at the $\Gamma$ point and PCBM is dispersionless, the interfacial electronic structure is most important at the $\Gamma$ point. We therefore only show and discuss the results for the interface at the $\Gamma$ point in this work, although the calculations do involve a $\mathbf{k}$-point sampling. We use the substrate screening approach \cite{Liu2019} as discussed in the Computational Methods section for interface electronic structure at the $GW$ level, which is compared to that of the isolated (P3HT)$_n$ and isolated PCBM. 

Fig. \ref{figure2} clearly shows that this interface is a ``type-II'' heterojunction. We align the vacuum level (the dotted line) of the isolated P3HT with the vacuum level of the interface on the P3HT side. Similarly, we align the vacuum level of the isolated PCBM with the vacuum level of the interface on the PCBM side, which is set to be zero in Fig. \ref{figure2}. The difference in the vacuum level across the interface is related to the interface dipole. Dashed lines are PBE energy levels and solid lines are $GW$ energy levels. At the interface, the fundamental gap (the black arrow) is between P3HT VBM and PCBM LUMO, and PBE significantly underestimates this gap compared to $GW$ (0.16 eV compared to 2.09 eV). Both the P3HT gap (the blue arrow) and the PCBM gap (the red arrow) are smaller at the interface than their isolated counterparts, suggesting that at such a molecule-semiconductor interface, both components could effectively screen the Coulomb interactions within the other component of the interface, resulting in gap renormalizations at both sides of the interface. This fact is qualitatively different from the molecule-metal interfaces \cite{Neaton2006,Thygesen2009}, where only the metallic substrate effectively screens the Coulomb interaction within the molecule and renormalizes its gap.

Importantly, what makes the level alignment picture at a molecule-semiconductor interface more complicated than that at a molecule-metal interface is the following: not only the fundamental gap of the interface and the adsorbate gap are important, but also how the adsorbate HOMO (LUMO) aligns with substrate VBM (CBM) is useful information. The latter is especially true when the interface is a ``type-II'' heterojunction such as the P3HT:PCBM considered here, because it is the driving force for the hole (electron) transfer barrier across the interface\cite{Ameri2013}. Because of the complex picture of the level alignment, $GW$ results are essential in quantitative understanding of the interfacial electronic structure, and the substrate screening approach developed in Ref. \citenum{Liu2019} proves to be a powerful tool for large interface systems, as we show in this paper.

\begin{table}[t!]
\caption{Main gaps at the $\Gamma$ point for the different interface systems studied in this work. The ``gap'' is the fundamental gap at the interface, between P3HT VBM and PCBM LUMO, denoted by the black arrow in Fig. \ref{figure2}. P3HT (PCBM) gap is the VBM-CBM (HOMO-LUMO) gap within the interface system, smaller than that of the isolated P3HT (PCBM) counterpart, which is denoted by the blue (red) arrow in Fig. \ref{figure2}. $\Delta_{\rm LL}$ ($\Delta_{\rm HH}$) is the gap between P3HT CBM (VBM) and PCBM LUMO (HOMO) and is numerically the difference between the ``gap'' and the P3HT (PCBM) gap, useful for understanding the electron (hole) transfer across a ``type-II'' heterojunction. All values are in eV and are from $GW$ results.}
\label{tab:table3}
\begin{tabular}{c|ccccc}
\hline
                & gap  & P3HT gap & PCBM gap & $\Delta_{\rm LL}$ & $\Delta_{\rm HH}$ \\ \hline
(P3HT)$_1$:PCBM & 2.09 & 2.66     & 3.65     & 0.57              & 1.56              \\
(P3HT)$_2$:PCBM & 1.79 & 2.04     & 3.48     & 0.25              & 1.69              \\
(P3HT)$_3$:PCBM & 1.69 & 1.78     & 3.42     & 0.09              & 1.73              \\
(P3HT)$_4$:PCBM & 1.59 & 1.64     & 3.37     & 0.05              & 1.78              \\ 
(P3HT)$_1$:PCBM:(P3HT)$_1$ & 1.30 & 2.06     & 2.52     & 0.76              & 1.22              \\ \hline
\end{tabular}
\end{table}

In Table \ref{tab:table3}, we summarize the level alignment results at the $\Gamma$ point for all systems we study in this work. One can see that for (P3HT)$_n$:PCBM, as the number of P3HT layers increases, the interface gap, P3HT gap, and PCBM gap all decrease, due to enhanced many-body screening effect. Compared to the isolated substrate results in Table \ref{tab:table1}, the P3HT gaps at the interface decrease by about 0.2-0.25 eV for all systems, because the adsorbate effect is similar across the series. Compared to the isolated adsorbate, the PCBM gap at the interface decreases by 0.53 eV, 0.70 eV, 0.76 eV, and 0.81 eV, respectively, across the series. This is because the many-body screening for the adsorbate becomes stronger when the substrate becomes thicker. The gap between P3HT CBM and PCBM LUMO ($\Delta_{\rm LL}$ in Table \ref{tab:table3}) becomes nearly zero for (P3HT)$_4$:PCBM, suggesting that a charge-transfer exciton across the interface is energetically similar to an exciton localized on the P3HT side. From the numerical trend, we also conclude that the electronic structure of the (P3HT)$_4$:PCBM interface is close to that of bulk P3HT:PCBM, i.e., (P3HT)$_n$:PCBM with $n\to \infty$. 

% one interface vs. two interfaces 

\begin{figure}[t!]
\centering
\includegraphics[width=3.5in]{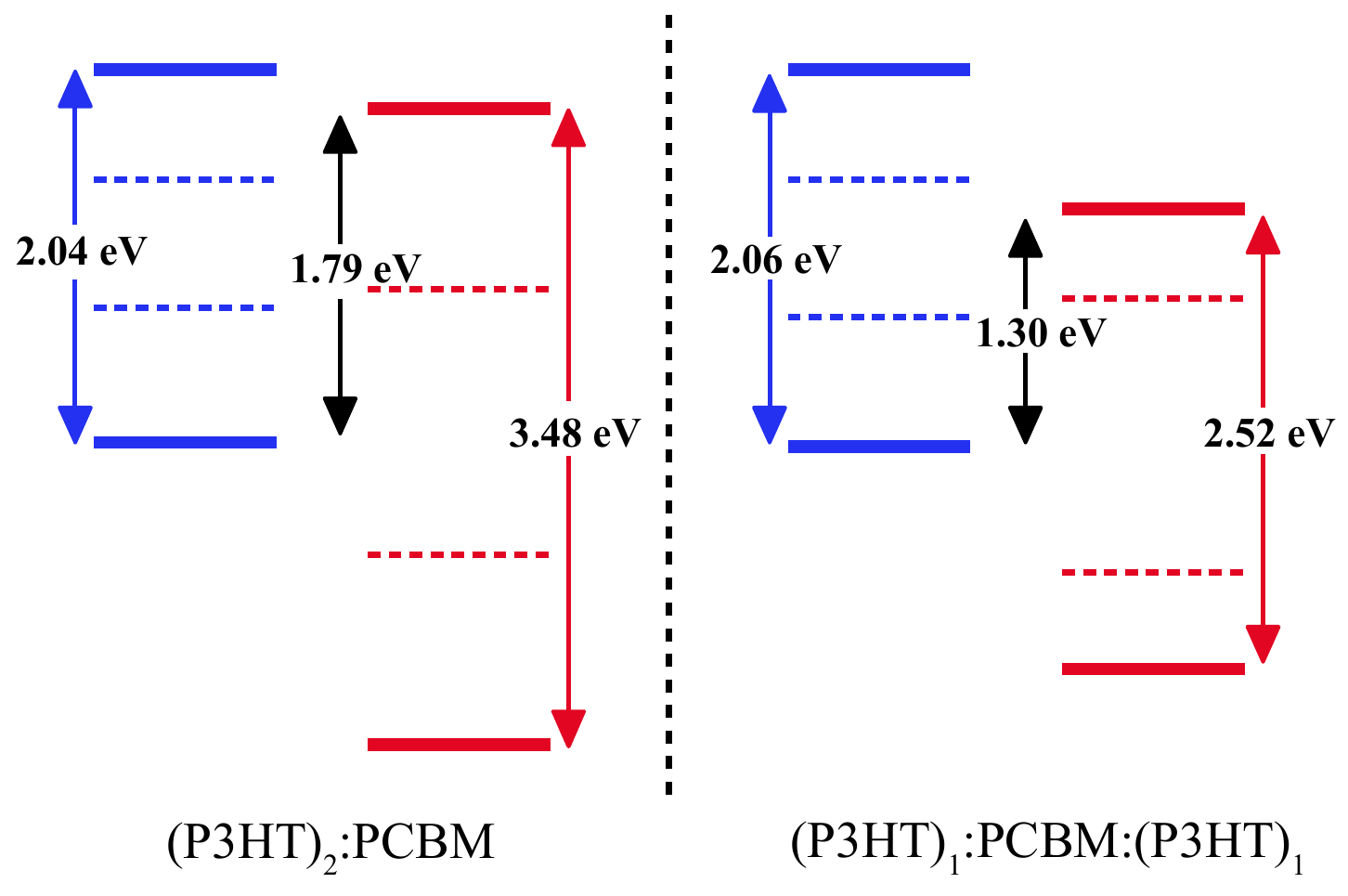}
\caption{Comparison of interface energy level alignment diagrams for (P3HT)$_2$:PCBM and (P3HT)$_1$:PCBM:(P3HT)$_1$. Blue (red) lines indicate the interface energy levels that are localized on P3HT (PCBM). Blue arrows indicate the VBM-CBM band gap of P3HT; red arrows indicate the HOMO-LUMO gap of PCBM; and the black arrows indicate the fundamental gaps of the interface systems. Solid lines are from $GW$ and dashed lines are from PBE.}
\label{compare}
\end{figure}

Comparing (P3HT)$_2$:PCBM and (P3HT)$_1$:PCBM:(P3HT)$_1$ in Table \ref{tab:table3} is especially instructive. In both systems there is one PCBM molecule and two layers of P3HT. The difference is that in the former, one P3HT:PCBM interface is formed [Fig. \ref{figure1}(b)]; and in the latter, two P3HT:PCBM interfaces are formed, one on either side of PCBM [Fig. \ref{figure1}(e)]. To better illustrate their difference in the interface electronic structure, we compare their level alignment diagrams in Fig. \ref{compare}. PBE results (dashed lines) are similar due to missing of long-range correlation effect, but $GW$ results (solid lines) differ significantly due to different dielectric environments around the PCBM molecule. The fundamental gaps (black arrows) of the two systems differ by about 0.5 eV. Furthermore, the PCBM gap is about 1 eV smaller in (P3HT)$_1$:PCBM:(P3HT)$_1$ than in (P3HT)$_2$:PCBM, due to enhanced many-body screening arising from interfaces on both sides of the PCBM. As a result, the difference between P3HT CBM and PCBM LUMO, $\Delta_{\rm LL}$, is 0.76 eV in (P3HT)$_1$:PCBM:(P3HT)$_1$, compared to 0.25 eV in (P3HT)$_2$:PCBM. $\Delta_{\rm LL}$ is usually reported to be about 0.7 eV in the literature \cite{Kroon2008,Li2012,Shih2013}, suggesting that interpenetrating structures similar to (P3HT)$_1$:PCBM:(P3HT)$_1$ are more prevalent in experiments, where the bulky PCBM molecule may even form more than two interfaces with P3HT. However, lower $\Delta_{\rm LL}$ could favor faster exciton dissociation since $\Delta_{\rm LL}$ is phenomenologically the charge transfer barrier across the interface\cite{BBCC04}. Our results here indicate that by carefully designing the placement of P3HT layers around PCBM, one can effectively tune the magnitude of $\Delta_{\rm LL}$. For instance, if the blend system can be fabricated such that the PCBM only forms an interface with P3HT on one side, with the other side filled with certain low-dielectric materials [mimicking the vacuum in our modeling of (P3HT)$_n$:PCBM], then the $\Delta_{\rm LL}$ can be significantly reduced compared to the regular interpenetrating network, a possibility that may be of interest for experimental verification.

\subsection{Comparison of Different $GW$-based Methods}
% different levels of GW theory                      

\begin{table}[h!]
\caption{Comparison of results from different $GW$-based approaches, for (P3HT)$_1$:PCBM. $\Delta_{\rm HOMO}$ is the $GW$ self-energy correction to the PBE HOMO of PCBM. $\Delta_{\rm LUMO}$ is the $GW$ self-energy correction to the PBE LUMO of PCBM. $\Delta\mbox{gap}=\Delta_{\rm HOMO}+\Delta_{\rm LUMO}$. For details, see text. All values are in eV.}
\label{tab:table4}
\begin{tabular}{l|ccc}
\hline
    & $\Delta_{\rm HOMO}$ & $\Delta_{\rm LUMO}$ & $\Delta$gap \\ \hline
(1) Direct $GW$ & 0.95                 & 1.17                 & 2.12        \\
(2) Substrate screening & 0.92                 & 1.21                 & 2.13        \\
(3) Projection $GW$ & 0.98                 & 1.15                 & 2.13        \\
(4) Embedding $GW$ & 0.96                 & 1.17                 & 2.13        \\ \hline
\end{tabular}
\end{table}

Finally, we briefly compare different levels of $GW$-based approximations for the adsorbate, using (P3HT)$_1$:PCBM as an example. The purpose is to validate new computational methods using complex systems such as those studied in this work. Such comparison will also benefit future computational method development. Table \ref{tab:table4} shows the results from four different $GW$-based approaches: (1) direct $GW$ calculation of the interface; (2) substrate screening $GW$, which is used for all (P3HT)$_n$:PCBM systems reported in Table \ref{tab:table3}; (3) projection $GW$ for the adsorbate, as developed in Refs. \citenum{Tamblyn2011} and \citenum{Chen2017}. This approach computes $\left<\phi_i\left|\Sigma\left[G^{\rm tot}W^{\rm tot}\right]\right|\phi_i\right>$, where $G^{\rm tot}$ and $W^{\rm tot}$ are the Green's function and screened Coulomb interaction for the combined interface, respectively; and (4) dielectric embedding $GW$, as developed in Ref. \citenum{Liu2020}. This approach computes $\left<\phi_i\left|\Sigma\left[G^{\rm ad}W^{\rm tot}\right]\right|\phi_i\right>$, where $G^{\rm ad}$ is the Green's function of the PCBM adsorbate and $W^{\rm tot}$ is the screened Coulomb interaction for the combined interface. In both (3) and (4), $\phi_i$ is either PCBM HOMO or LUMO and the self-energy calculation is performed in a simulation cell that is the same size as the interface but contains just the PCBM. Ref. \citenum{Liu2019} showed that (2) and (1) yield very similar results and Ref. \citenum{Liu2020} showed that (4) and (3) yield very similar results, for molecule-metal interfaces. Here we demonstrate that the same conclusion holds for a complex molecule-semiconductor interface, and all methods agree within about 0.05 eV for $GW$ corrections to individual PBE energy levels of the adsorbate. These results provide additional evidence for the reliability of the methods that we developed in Refs. \citenum{Liu2019,Liu2020}, and serve as the first step for future method development on efficient BSE calculations of the interface.

\section{Conclusions}
In this work, we used $GW$-based approaches to quantitatively study the electronic structure of (P3HT)$_n$:PCBM ($n$ = 1, 2, 3, and 4) and (P3HT)$_1$:PCBM:(P3HT)$_1$ heterostructures. We focused on how the number of P3HT layers affects the energy level alignment at the interface, as well as a direct comparison between (P3HT)$_2$:PCBM and (P3HT)$_1$:PCBM:(P3HT)$_1$. We quantitatively determined the energy level alignment of the ``type-II'' heterostructures, and showed that the electronic structure can be effectively tuned by the thickness of P3HT layers and different placements of the P3HT layers around PCBM. Our results showed that the substrate and the adsorbate simultaneously screen the Coulomb interaction within each other, resulting in smaller gaps at the interface compared to their respective isolated phases. We used the substrate screening $GW$ approach for the heterogeneous interfaces, which is as accurate as the direct $GW$ but is more computationally affordable, showing that this approach is promising in treating complex interface systems. Our results provide computational insights into understanding of such complex heterostructures by unveiling unambiguous structure-property relationships and highlighting many-body screening effects at the interface. 

\section{Supporting Information Description}
Tables showing the identification of interface orbitals as P3HT or PCBM frontier orbitals, as well as figures showing the energy level diagrams for (P3HT)$_n$:PCBM with $n \geqslant 2$ and (P3HT)$_1$:PCBM:(P3HT)$_1$. 

\section{Acknowledgements}
Z.-F.L. thanks Wayne State University for generous start-up funds. This research was supported by a grant from the United States-Israel Binational Science Foundation (BSF), Jerusalem, Israel, awarded to Z.-F.L. and S.R.-A. (Grant Number 2018113). The authors thankfully acknowledge PRACE for awarding us access to the computer resources at MareNostrum and the technical support provided by Barcelona Supercomputing Center.

\bibliography{P3ht}

\providecommand{\latin}[1]{#1}
\makeatletter
\providecommand{\doi}
  {\begingroup\let\do\@makeother\dospecials
  \catcode`\{=1 \catcode`\}=2 \doi@aux}
\providecommand{\doi@aux}[1]{\endgroup\texttt{#1}}
\makeatother
\providecommand*\mcitethebibliography{\thebibliography}
\csname @ifundefined\endcsname{endmcitethebibliography}
  {\let\endmcitethebibliography\endthebibliography}{}
\begin{mcitethebibliography}{66}
\providecommand*\natexlab[1]{#1}
\providecommand*\mciteSetBstSublistMode[1]{}
\providecommand*\mciteSetBstMaxWidthForm[2]{}
\providecommand*\mciteBstWouldAddEndPuncttrue
  {\def\EndOfBibitem{\unskip.}}
\providecommand*\mciteBstWouldAddEndPunctfalse
  {\let\EndOfBibitem\relax}
\providecommand*\mciteSetBstMidEndSepPunct[3]{}
\providecommand*\mciteSetBstSublistLabelBeginEnd[3]{}
\providecommand*\EndOfBibitem{}
\mciteSetBstSublistMode{f}
\mciteSetBstMaxWidthForm{subitem}{(\alph{mcitesubitemcount})}
\mciteSetBstSublistLabelBeginEnd
  {\mcitemaxwidthsubitemform\space}
  {\relax}
  {\relax}

\bibitem[Kaltenbrunner \latin{et~al.}(2012)Kaltenbrunner, White, G{\l}owacki,
  Sekitani, Someya, Sariciftci, and Bauer]{Kaltenbrunner2012}
Kaltenbrunner,~M.; White,~M.~S.; G{\l}owacki,~E.~D.; Sekitani,~T.; Someya,~T.;
  Sariciftci,~N.~S.; Bauer,~S. Ultrathin and lightweight organic solar cells
  with high flexibility. \emph{Nat. Commun.} \textbf{2012}, \emph{3}, 770\relax
\mciteBstWouldAddEndPuncttrue
\mciteSetBstMidEndSepPunct{\mcitedefaultmidpunct}
{\mcitedefaultendpunct}{\mcitedefaultseppunct}\relax
\EndOfBibitem
\bibitem[Dennler \latin{et~al.}(2009)Dennler, Scharber, and
  Brabec]{Dennler2009}
Dennler,~G.; Scharber,~M.~C.; Brabec,~C.~J. Polymer-Fullerene
  Bulk-Heterojunction Solar Cells. \emph{Adv. Mater.} \textbf{2009}, \emph{21},
  1323--1338\relax
\mciteBstWouldAddEndPuncttrue
\mciteSetBstMidEndSepPunct{\mcitedefaultmidpunct}
{\mcitedefaultendpunct}{\mcitedefaultseppunct}\relax
\EndOfBibitem
\bibitem[S{\o}ndergaard \latin{et~al.}(2012)S{\o}ndergaard, H{\"o}sel, Angmo,
  Larsen-Olsen, and Krebs]{Soendergaard2012}
S{\o}ndergaard,~R.; H{\"o}sel,~M.; Angmo,~D.; Larsen-Olsen,~T.~T.; Krebs,~F.~C.
  Roll-to-roll fabrication of polymer solar cells. \emph{Mater. Today}
  \textbf{2012}, \emph{15}, 36--49\relax
\mciteBstWouldAddEndPuncttrue
\mciteSetBstMidEndSepPunct{\mcitedefaultmidpunct}
{\mcitedefaultendpunct}{\mcitedefaultseppunct}\relax
\EndOfBibitem
\bibitem[Krebs \latin{et~al.}(2010)Krebs, Tromholt, and
  J{\o}rgensen]{Krebs2010}
Krebs,~F.~C.; Tromholt,~T.; J{\o}rgensen,~M. Upscaling of polymer solar cell
  fabrication using full roll-to-roll processing. \emph{Nanoscale}
  \textbf{2010}, \emph{2}, 873--886\relax
\mciteBstWouldAddEndPuncttrue
\mciteSetBstMidEndSepPunct{\mcitedefaultmidpunct}
{\mcitedefaultendpunct}{\mcitedefaultseppunct}\relax
\EndOfBibitem
\bibitem[Hains \latin{et~al.}(2010)Hains, Ramanan, Irwin, Liu, Wasielewski, and
  Marks]{Hains2010}
Hains,~A.~W.; Ramanan,~C.; Irwin,~M.~D.; Liu,~J.; Wasielewski,~M.~R.;
  Marks,~T.~J. Designed Bithiophene-Based Interfacial Layer for High-Efficiency
  Bulk-Heterojunction Organic Photovoltaic Cells. Importance of Interfacial
  Energy Level Matching. \emph{ACS Appl. Mater. Interfaces} \textbf{2010},
  \emph{2}, 175--185\relax
\mciteBstWouldAddEndPuncttrue
\mciteSetBstMidEndSepPunct{\mcitedefaultmidpunct}
{\mcitedefaultendpunct}{\mcitedefaultseppunct}\relax
\EndOfBibitem
\bibitem[Kobori \latin{et~al.}({2013})Kobori, Noji, and
  Tsuganezawa]{Kobori2013}
Kobori,~Y.; Noji,~R.; Tsuganezawa,~S. {Initial Molecular Photocurrent:
  Nanostructure and Motion of Weakly Bound Charge-Separated State in Organic
  Photovoltaic Interface}. \emph{{J. Phys. Chem. C}} \textbf{{2013}},
  \emph{{117}}, {1589--1599}\relax
\mciteBstWouldAddEndPuncttrue
\mciteSetBstMidEndSepPunct{\mcitedefaultmidpunct}
{\mcitedefaultendpunct}{\mcitedefaultseppunct}\relax
\EndOfBibitem
\bibitem[Fazzi \latin{et~al.}({2017})Fazzi, Barbatti, and Thiel]{Fazzi2017}
Fazzi,~D.; Barbatti,~M.; Thiel,~W. {Hot and Cold Charge-Transfer Mechanisms in
  Organic Photovoltaics: Insights into the Excited States of Donor/Acceptor
  Interfaces}. \emph{{J. Phys. Chem. Lett.}} \textbf{{2017}}, \emph{{8}},
  {4727--4734}\relax
\mciteBstWouldAddEndPuncttrue
\mciteSetBstMidEndSepPunct{\mcitedefaultmidpunct}
{\mcitedefaultendpunct}{\mcitedefaultseppunct}\relax
\EndOfBibitem
\bibitem[Menichetti \latin{et~al.}({2017})Menichetti, Colle, and
  Grosso]{Menichetti2017}
Menichetti,~G.; Colle,~R.; Grosso,~G. {Strain Modulation of Band Offsets at the
  PCBM/P3HT Heterointerface}. \emph{{J. Phys. Chem. C}} \textbf{{2017}},
  \emph{{121}}, {13707--13716}\relax
\mciteBstWouldAddEndPuncttrue
\mciteSetBstMidEndSepPunct{\mcitedefaultmidpunct}
{\mcitedefaultendpunct}{\mcitedefaultseppunct}\relax
\EndOfBibitem
\bibitem[Krebs \latin{et~al.}(2009)Krebs, Gevorgyan, and Alstrup]{Krebs2009}
Krebs,~F.~C.; Gevorgyan,~S.~A.; Alstrup,~J. A roll-to-roll process to flexible
  polymer solar cells: model studies{,} manufacture and operational stability
  studies. \emph{J. Mater. Chem.} \textbf{2009}, \emph{19}, 5442--5451\relax
\mciteBstWouldAddEndPuncttrue
\mciteSetBstMidEndSepPunct{\mcitedefaultmidpunct}
{\mcitedefaultendpunct}{\mcitedefaultseppunct}\relax
\EndOfBibitem
\bibitem[Gonz{\'a}lez \latin{et~al.}(2015)Gonz{\'a}lez, K{\"o}rstgens, Yao,
  Song, Santoro, Roth, and M{\"u}ller-Buschbaum]{Gonzalez2015}
Gonz{\'a}lez,~D.~M.; K{\"o}rstgens,~V.; Yao,~Y.; Song,~L.; Santoro,~G.;
  Roth,~S.~V.; M{\"u}ller-Buschbaum,~P. Improved Power Conversion Efficiency of
  P3HT:PCBM Organic Solar Cells by Strong Spin--Orbit Coupling-Induced Delayed
  Fluorescence. \emph{Adv. Energy Mater.} \textbf{2015}, \emph{5},
  1401770\relax
\mciteBstWouldAddEndPuncttrue
\mciteSetBstMidEndSepPunct{\mcitedefaultmidpunct}
{\mcitedefaultendpunct}{\mcitedefaultseppunct}\relax
\EndOfBibitem
\bibitem[Liu and Troisi(2011)Liu, and Troisi]{Liu2011}
Liu,~T.; Troisi,~A. Absolute Rate of Charge Separation and Recombination in a
  Molecular Model of the P3HT/PCBM Interface. \emph{J. Phys. Chem. C}
  \textbf{2011}, \emph{115}, 2406--2415\relax
\mciteBstWouldAddEndPuncttrue
\mciteSetBstMidEndSepPunct{\mcitedefaultmidpunct}
{\mcitedefaultendpunct}{\mcitedefaultseppunct}\relax
\EndOfBibitem
\bibitem[D'Avino \latin{et~al.}(2016)D'Avino, Muccioli, Olivier, and
  Beljonne]{DMOB16}
D'Avino,~G.; Muccioli,~L.; Olivier,~Y.; Beljonne,~D. Charge Separation and
  Recombination at Polymer-Fullerene Heterojunctions: Delocalization and
  Hybridization Effects. \emph{J. Phys. Chem. Lett.} \textbf{2016}, \emph{7},
  536--540\relax
\mciteBstWouldAddEndPuncttrue
\mciteSetBstMidEndSepPunct{\mcitedefaultmidpunct}
{\mcitedefaultendpunct}{\mcitedefaultseppunct}\relax
\EndOfBibitem
\bibitem[Jakowetz \latin{et~al.}(2017)Jakowetz, B{\"o}hm, Sadhanala, Huettner,
  Rao, and Friend]{Jakowetz2017}
Jakowetz,~A.~C.; B{\"o}hm,~M.~L.; Sadhanala,~A.; Huettner,~S.; Rao,~A.;
  Friend,~R.~H. Visualizing excitations at buried heterojunctions in organic
  semiconductor blends. \emph{Nat. Mater.} \textbf{2017}, \emph{16},
  551--557\relax
\mciteBstWouldAddEndPuncttrue
\mciteSetBstMidEndSepPunct{\mcitedefaultmidpunct}
{\mcitedefaultendpunct}{\mcitedefaultseppunct}\relax
\EndOfBibitem
\bibitem[Yin and Dadmun(2011)Yin, and Dadmun]{Yin2011}
Yin,~W.; Dadmun,~M. A New Model for the Morphology of P3HT/PCBM Organic
  Photovoltaics from Small-Angle Neutron Scattering: Rivers and Streams.
  \emph{ACS Nano} \textbf{2011}, \emph{5}, 4756--4768\relax
\mciteBstWouldAddEndPuncttrue
\mciteSetBstMidEndSepPunct{\mcitedefaultmidpunct}
{\mcitedefaultendpunct}{\mcitedefaultseppunct}\relax
\EndOfBibitem
\bibitem[Vandewal \latin{et~al.}(2014)Vandewal, Albrecht, Hoke, Graham, Widmer,
  Douglas, Schubert, Mateker, Bloking, Burkhard, Sellinger, Frechet, Amassian,
  Riede, McGehee, Neher, and Salleo]{Vandewal2014}
Vandewal,~K.; Albrecht,~S.; Hoke,~E.~T.; Graham,~K.~R.; Widmer,~J.;
  Douglas,~J.~D.; Schubert,~M.; Mateker,~W.~R.; Bloking,~J.~T.;
  Burkhard,~G.~F.; Sellinger,~A.; Frechet,~J. M.~J.; Amassian,~A.;
  Riede,~M.~K.; McGehee,~M.~D.; Neher,~D.; Salleo,~A. Efficient charge
  generation by relaxed charge-transfer states at organic interfaces.
  \emph{Nat. Mater.} \textbf{2014}, \emph{13}, 63--68\relax
\mciteBstWouldAddEndPuncttrue
\mciteSetBstMidEndSepPunct{\mcitedefaultmidpunct}
{\mcitedefaultendpunct}{\mcitedefaultseppunct}\relax
\EndOfBibitem
\bibitem[Deotare \latin{et~al.}(2015)Deotare, Chang, Hontz, Congreve, Shi,
  Reusswig, Modtland, Bahlke, Lee, Willard, Bulovic, Van~Voorhis, and
  Baldo]{Deotare2015}
Deotare,~P.~B.; Chang,~W.; Hontz,~E.; Congreve,~D.~N.; Shi,~L.;
  Reusswig,~P.~D.; Modtland,~B.; Bahlke,~M.~E.; Lee,~C.~K.; Willard,~A.~P.;
  Bulovic,~V.; Van~Voorhis,~T.; Baldo,~M.~A. Nanoscale transport of
  charge-transfer states in organic donor-acceptor blends. \emph{Nat. Mater.}
  \textbf{2015}, \emph{14}, 1130--1134\relax
\mciteBstWouldAddEndPuncttrue
\mciteSetBstMidEndSepPunct{\mcitedefaultmidpunct}
{\mcitedefaultendpunct}{\mcitedefaultseppunct}\relax
\EndOfBibitem
\bibitem[Bittner and Silva(2014)Bittner, and Silva]{Bittner2014}
Bittner,~E.~R.; Silva,~C. Noise-induced quantum coherence drives photo-carrier
  generation dynamics at polymeric semiconductor heterojunctions. \emph{Nat.
  Commun.} \textbf{2014}, \emph{5}, 3119\relax
\mciteBstWouldAddEndPuncttrue
\mciteSetBstMidEndSepPunct{\mcitedefaultmidpunct}
{\mcitedefaultendpunct}{\mcitedefaultseppunct}\relax
\EndOfBibitem
\bibitem[Grancini \latin{et~al.}(2011)Grancini, Polli, Fazzi,
  Cabanillas-Gonzalez, Cerullo, and Lanzani]{Grancini2011}
Grancini,~G.; Polli,~D.; Fazzi,~D.; Cabanillas-Gonzalez,~J.; Cerullo,~G.;
  Lanzani,~G. Transient Absorption Imaging of P3HT:PCBM Photovoltaic Blend:
  Evidence For Interfacial Charge Transfer State. \emph{J. Phys. Chem. Lett.}
  \textbf{2011}, \emph{2}, 1099--1105\relax
\mciteBstWouldAddEndPuncttrue
\mciteSetBstMidEndSepPunct{\mcitedefaultmidpunct}
{\mcitedefaultendpunct}{\mcitedefaultseppunct}\relax
\EndOfBibitem
\bibitem[Mothy \latin{et~al.}(2012)Mothy, Guillaume, Id{\'e}, Castet, Ducasse,
  Cornil, and Beljonne]{Mothy2012}
Mothy,~S.; Guillaume,~M.; Id{\'e},~J.; Castet,~F.; Ducasse,~L.; Cornil,~J.;
  Beljonne,~D. Tuning the Interfacial Electronic Structure at Organic
  Heterojunctions by Chemical Design. \emph{J. Phys. Chem. Lett.}
  \textbf{2012}, \emph{3}, 2374--2378\relax
\mciteBstWouldAddEndPuncttrue
\mciteSetBstMidEndSepPunct{\mcitedefaultmidpunct}
{\mcitedefaultendpunct}{\mcitedefaultseppunct}\relax
\EndOfBibitem
\bibitem[To and Adams(2014)To, and Adams]{To2014}
To,~T.~T.; Adams,~S. Modelling of P3HT:PCBM interface using coarse-grained
  forcefield derived from accurate atomistic forcefield. \emph{Phys. Chem.
  Chem. Phys.} \textbf{2014}, \emph{16}, 4653--4663\relax
\mciteBstWouldAddEndPuncttrue
\mciteSetBstMidEndSepPunct{\mcitedefaultmidpunct}
{\mcitedefaultendpunct}{\mcitedefaultseppunct}\relax
\EndOfBibitem
\bibitem[Kanai and Grossman(2007)Kanai, and Grossman]{Kanai2007}
Kanai,~Y.; Grossman,~J.~C. Insights on Interfacial Charge Transfer Across
  P3HT/Fullerene Photovoltaic Heterojunction from Ab Initio Calculations.
  \emph{Nano Lett.} \textbf{2007}, \emph{7}, 1967--1972\relax
\mciteBstWouldAddEndPuncttrue
\mciteSetBstMidEndSepPunct{\mcitedefaultmidpunct}
{\mcitedefaultendpunct}{\mcitedefaultseppunct}\relax
\EndOfBibitem
\bibitem[{Hohenberg} and {Kohn}(1964){Hohenberg}, and {Kohn}]{Hohenberg1964}
{Hohenberg},~P.; {Kohn},~W. {Inhomogeneous Electron Gas}. \emph{Phys. Rev.}
  \textbf{1964}, \emph{136}, 864--871\relax
\mciteBstWouldAddEndPuncttrue
\mciteSetBstMidEndSepPunct{\mcitedefaultmidpunct}
{\mcitedefaultendpunct}{\mcitedefaultseppunct}\relax
\EndOfBibitem
\bibitem[Kohn and Sham(1965)Kohn, and Sham]{Kohn1965}
Kohn,~W.; Sham,~L.~J. Self-Consistent Equations Including Exchange and
  Correlation Effects. \emph{Phys. Rev.} \textbf{1965}, \emph{140},
  A1133--A1138\relax
\mciteBstWouldAddEndPuncttrue
\mciteSetBstMidEndSepPunct{\mcitedefaultmidpunct}
{\mcitedefaultendpunct}{\mcitedefaultseppunct}\relax
\EndOfBibitem
\bibitem[Koch and Holthausen(2002)Koch, and Holthausen]{2001}
Koch,~W.; Holthausen,~M.~C. \emph{A Chemist's Guide to Density Functional
  Theory}; 2nd ed. Wiley-VCH (Weinheim), 2002\relax
\mciteBstWouldAddEndPuncttrue
\mciteSetBstMidEndSepPunct{\mcitedefaultmidpunct}
{\mcitedefaultendpunct}{\mcitedefaultseppunct}\relax
\EndOfBibitem
\bibitem[K{\"u}mmel and Kronik(2008)K{\"u}mmel, and Kronik]{Kuemmel2008}
K{\"u}mmel,~S.; Kronik,~L. Orbital-dependent density functionals: Theory and
  applications. \emph{Rev. Mod. Phys.} \textbf{2008}, \emph{80}, 3--60\relax
\mciteBstWouldAddEndPuncttrue
\mciteSetBstMidEndSepPunct{\mcitedefaultmidpunct}
{\mcitedefaultendpunct}{\mcitedefaultseppunct}\relax
\EndOfBibitem
\bibitem[Perdew and Levy(1983)Perdew, and Levy]{Perdew1983}
Perdew,~J.~P.; Levy,~M. Physical Content of the Exact Kohn-Sham Orbital
  Energies: Band Gaps and Derivative Discontinuities. \emph{Phys. Rev. B}
  \textbf{1983}, \emph{51}, 1884--1887\relax
\mciteBstWouldAddEndPuncttrue
\mciteSetBstMidEndSepPunct{\mcitedefaultmidpunct}
{\mcitedefaultendpunct}{\mcitedefaultseppunct}\relax
\EndOfBibitem
\bibitem[Yang \latin{et~al.}(2012)Yang, Cohen, and Mori-S{\'a}nchez]{Yang2012}
Yang,~W.; Cohen,~A.~J.; Mori-S{\'a}nchez,~P. Derivative discontinuity, bandgap
  and lowest unoccupied molecular orbital in density functional theory.
  \emph{J. Chem. Phys.} \textbf{2012}, \emph{136}, 204111\relax
\mciteBstWouldAddEndPuncttrue
\mciteSetBstMidEndSepPunct{\mcitedefaultmidpunct}
{\mcitedefaultendpunct}{\mcitedefaultseppunct}\relax
\EndOfBibitem
\bibitem[Neaton \latin{et~al.}(2006)Neaton, Hybertsen, and Louie]{Neaton2006}
Neaton,~J.~B.; Hybertsen,~M.~S.; Louie,~S.~G. Renormalization of Molecular
  Electronic Levels at Metal-Molecule Interfaces. \emph{Phys. Rev. Lett.}
  \textbf{2006}, \emph{97}, 216405\relax
\mciteBstWouldAddEndPuncttrue
\mciteSetBstMidEndSepPunct{\mcitedefaultmidpunct}
{\mcitedefaultendpunct}{\mcitedefaultseppunct}\relax
\EndOfBibitem
\bibitem[Hedin(1965)]{Hedin1965}
Hedin,~L. New Method for Calculating the One-Particle Green's Function with
  Application to the Electron-Gas Problem. \emph{Phys. Rev.} \textbf{1965},
  \emph{139}, A796--A823\relax
\mciteBstWouldAddEndPuncttrue
\mciteSetBstMidEndSepPunct{\mcitedefaultmidpunct}
{\mcitedefaultendpunct}{\mcitedefaultseppunct}\relax
\EndOfBibitem
\bibitem[Hybertsen and Louie(1986)Hybertsen, and Louie]{Hybertsen1986}
Hybertsen,~M.~S.; Louie,~S.~G. Electron correlation in semiconductors and
  insulators: Band gaps and quasiparticle energies. \emph{Phys. Rev. B}
  \textbf{1986}, \emph{34}, 5390--5413\relax
\mciteBstWouldAddEndPuncttrue
\mciteSetBstMidEndSepPunct{\mcitedefaultmidpunct}
{\mcitedefaultendpunct}{\mcitedefaultseppunct}\relax
\EndOfBibitem
\bibitem[Li \latin{et~al.}(2014)Li, Kontsevoi, and Freeman]{Li2014}
Li,~L.-H.; Kontsevoi,~O.~Y.; Freeman,~A.~J. Orientation-Dependent Electronic
  Structures and Optical Properties of the P3HT:PCBM Interface: A
  First-Principles GW-BSE Study. \emph{J. Phys. Chem. C} \textbf{2014},
  \emph{118}, 10263--10270\relax
\mciteBstWouldAddEndPuncttrue
\mciteSetBstMidEndSepPunct{\mcitedefaultmidpunct}
{\mcitedefaultendpunct}{\mcitedefaultseppunct}\relax
\EndOfBibitem
\bibitem[Baumeier \latin{et~al.}(2012)Baumeier, Andrienko, and
  Rohlfing]{Baumeier2012}
Baumeier,~B.; Andrienko,~D.; Rohlfing,~M. Frenkel and Charge-Transfer
  Excitations in Donor-acceptor Complexes from Many-Body Green's Functions
  Theory. \emph{J. Chem. Theory Comput.} \textbf{2012}, \emph{8},
  2790--2795\relax
\mciteBstWouldAddEndPuncttrue
\mciteSetBstMidEndSepPunct{\mcitedefaultmidpunct}
{\mcitedefaultendpunct}{\mcitedefaultseppunct}\relax
\EndOfBibitem
\bibitem[Duchemin and Blase(2013)Duchemin, and Blase]{Duchemin2013}
Duchemin,~I.; Blase,~X. Resonant hot charge-transfer excitations in
  fullerene-porphyrin complexes: Many-body Bethe-Salpeter study. \emph{Phys.
  Rev. B} \textbf{2013}, \emph{87}, 245412\relax
\mciteBstWouldAddEndPuncttrue
\mciteSetBstMidEndSepPunct{\mcitedefaultmidpunct}
{\mcitedefaultendpunct}{\mcitedefaultseppunct}\relax
\EndOfBibitem
\bibitem[Thygesen and Rubio(2009)Thygesen, and Rubio]{Thygesen2009}
Thygesen,~K.~S.; Rubio,~A. Renormalization of Molecular Quasiparticle Levels at
  Metal-Molecule Interfaces: Trends across Binding Regimes. \emph{Phys. Rev.
  Lett.} \textbf{2009}, \emph{102}, 046802\relax
\mciteBstWouldAddEndPuncttrue
\mciteSetBstMidEndSepPunct{\mcitedefaultmidpunct}
{\mcitedefaultendpunct}{\mcitedefaultseppunct}\relax
\EndOfBibitem
\bibitem[Inkson(1971)]{Inkson1971}
Inkson,~J.~C. The electrostatic image potential in metal semiconductor
  junctions. \emph{J. Phys. C: Solid St. Phys.} \textbf{1971}, \emph{4},
  591--597\relax
\mciteBstWouldAddEndPuncttrue
\mciteSetBstMidEndSepPunct{\mcitedefaultmidpunct}
{\mcitedefaultendpunct}{\mcitedefaultseppunct}\relax
\EndOfBibitem
\bibitem[Lang and Kohn(1973)Lang, and Kohn]{Lang1973}
Lang,~N.~D.; Kohn,~W. Theory of Metal Surfaces: Induced Surface Charge and
  Image Potential. \emph{Phys. Rev. B} \textbf{1973}, \emph{7},
  3541--3550\relax
\mciteBstWouldAddEndPuncttrue
\mciteSetBstMidEndSepPunct{\mcitedefaultmidpunct}
{\mcitedefaultendpunct}{\mcitedefaultseppunct}\relax
\EndOfBibitem
\bibitem[Liu \latin{et~al.}(2019)Liu, da~Jornada, Louie, and Neaton]{Liu2019}
Liu,~Z.-F.; da~Jornada,~F.~H.; Louie,~S.~G.; Neaton,~J.~B. Accelerating
  GW-Based Energy Level Alignment Calculations for Molecule-Metal Interfaces
  Using a Substrate Screening Approach. \emph{J. Chem. Theory Comput.}
  \textbf{2019}, \emph{15}, 4218--4227\relax
\mciteBstWouldAddEndPuncttrue
\mciteSetBstMidEndSepPunct{\mcitedefaultmidpunct}
{\mcitedefaultendpunct}{\mcitedefaultseppunct}\relax
\EndOfBibitem
\bibitem[Ugeda \latin{et~al.}(2014)Ugeda, Bradley, Shi, da~Jornada, Zhang, Qiu,
  Ruan, Mo, Hussain, Shen, Wang, Louie, and Crommie]{Ugeda2014}
Ugeda,~M.~M.; Bradley,~A.~J.; Shi,~S.-F.; da~Jornada,~F.~H.; Zhang,~Y.;
  Qiu,~D.~Y.; Ruan,~W.; Mo,~S.-K.; Hussain,~Z.; Shen,~Z.-X.; Wang,~F.;
  Louie,~S.~G.; Crommie,~M.~F. Giant bandgap renormalization and excitonic
  effects in a monolayer transition metal dichalcogenide semiconductor.
  \emph{Nat. Mater.} \textbf{2014}, \emph{13}, 1091--1095\relax
\mciteBstWouldAddEndPuncttrue
\mciteSetBstMidEndSepPunct{\mcitedefaultmidpunct}
{\mcitedefaultendpunct}{\mcitedefaultseppunct}\relax
\EndOfBibitem
\bibitem[Xuan \latin{et~al.}(2019)Xuan, Chen, and Quek]{Xuan2019}
Xuan,~F.; Chen,~Y.; Quek,~S.~Y. Quasiparticle Levels at Large Interface Systems
  from Many-Body Perturbation Theory: The XAF-GW Method. \emph{J. Chem. Theory
  Comput.} \textbf{2019}, \emph{15}, 3824--3835\relax
\mciteBstWouldAddEndPuncttrue
\mciteSetBstMidEndSepPunct{\mcitedefaultmidpunct}
{\mcitedefaultendpunct}{\mcitedefaultseppunct}\relax
\EndOfBibitem
\bibitem[Momma and Izumi(2008)Momma, and Izumi]{Momma2008}
Momma,~K.; Izumi,~F. VESTA: a three-dimensional visualization system for
  electronic and structural analysis. \emph{J. Appl. Cryst.} \textbf{2008},
  \emph{41}, 653--658\relax
\mciteBstWouldAddEndPuncttrue
\mciteSetBstMidEndSepPunct{\mcitedefaultmidpunct}
{\mcitedefaultendpunct}{\mcitedefaultseppunct}\relax
\EndOfBibitem
\bibitem[Giannozzi \latin{et~al.}(2009)Giannozzi, Baroni, Bonini, Calandra,
  Car, Cavazzoni, Ceresoli, Chiarotti, Cococcioni, Dabo, {Dal Corso},
  de~Gironcoli, Fabris, Fratesi, Gebauer, Gerstmann, Gougoussis, Kokalj,
  Lazzeri, Martin-Samos, Marzari, Mauri, Mazzarello, Paolini, Pasquarello,
  Paulatto, Sbraccia, Scandolo, Sclauzero, Seitsonen, Smogunov, Umari, and
  Wentzcovitch]{Giannozzi2009}
Giannozzi,~P.; Baroni,~S.; Bonini,~N.; Calandra,~M.; Car,~R.; Cavazzoni,~C.;
  Ceresoli,~D.; Chiarotti,~G.~L.; Cococcioni,~M.; Dabo,~I.; {Dal Corso},~A.;
  de~Gironcoli,~S.; Fabris,~S.; Fratesi,~G.; Gebauer,~R.; Gerstmann,~U.;
  Gougoussis,~C.; Kokalj,~A.; Lazzeri,~M.; Martin-Samos,~L.; Marzari,~N.;
  Mauri,~F.; Mazzarello,~R.; Paolini,~S.; Pasquarello,~A.; Paulatto,~L.;
  Sbraccia,~C.; Scandolo,~S.; Sclauzero,~G.; Seitsonen,~A.~P.; Smogunov,~A.;
  Umari,~P.; Wentzcovitch,~R.~M. QUANTUM ESPRESSO: A modular and open-source
  software project for quantum simulations of materials. \emph{J. Phys.:
  Condens. Matter} \textbf{2009}, \emph{21}, 395502\relax
\mciteBstWouldAddEndPuncttrue
\mciteSetBstMidEndSepPunct{\mcitedefaultmidpunct}
{\mcitedefaultendpunct}{\mcitedefaultseppunct}\relax
\EndOfBibitem
\bibitem[Schlipf and Gygi(2015)Schlipf, and Gygi]{Schlipf2015}
Schlipf,~M.; Gygi,~F. Optimization algorithm for the generation of ONCV
  pseudopotentials. \emph{Comput. Phys. Commun.} \textbf{2015}, \emph{196},
  36--44\relax
\mciteBstWouldAddEndPuncttrue
\mciteSetBstMidEndSepPunct{\mcitedefaultmidpunct}
{\mcitedefaultendpunct}{\mcitedefaultseppunct}\relax
\EndOfBibitem
\bibitem[Hamann(2013)]{Hamann2013}
Hamann,~D.~R. Optimized norm-conserving Vanderbilt pseudopotentials.
  \emph{Phys. Rev. B} \textbf{2013}, \emph{88}, 085117\relax
\mciteBstWouldAddEndPuncttrue
\mciteSetBstMidEndSepPunct{\mcitedefaultmidpunct}
{\mcitedefaultendpunct}{\mcitedefaultseppunct}\relax
\EndOfBibitem
\bibitem[Perdew \latin{et~al.}(1996)Perdew, Burke, and Ernzerhof]{Perdew1996}
Perdew,~J.~P.; Burke,~K.; Ernzerhof,~M. Generalized Gradient Approximation Made
  Simple. \emph{Phys. Rev. Lett.} \textbf{1996}, \emph{77}, 3865--3868\relax
\mciteBstWouldAddEndPuncttrue
\mciteSetBstMidEndSepPunct{\mcitedefaultmidpunct}
{\mcitedefaultendpunct}{\mcitedefaultseppunct}\relax
\EndOfBibitem
\bibitem[Berland and Hyldgaard(2014)Berland, and Hyldgaard]{Berland2014}
Berland,~K.; Hyldgaard,~P. Exchange functional that tests the robustness of the
  plasmon description of the van der Waals density functional. \emph{Phys. Rev.
  B} \textbf{2014}, \emph{89}, 035412\relax
\mciteBstWouldAddEndPuncttrue
\mciteSetBstMidEndSepPunct{\mcitedefaultmidpunct}
{\mcitedefaultendpunct}{\mcitedefaultseppunct}\relax
\EndOfBibitem
\bibitem[{Deslippe} \latin{et~al.}(2012){Deslippe}, {Samsonidze}, {Strubbe},
  {Jain}, {Cohen}, and {Louie}]{Deslippe2012}
{Deslippe},~J.; {Samsonidze},~G.; {Strubbe},~D.~A.; {Jain},~M.; {Cohen},~M.~L.;
  {Louie},~S.~G. {BerkeleyGW: A massively parallel computer package for the
  calculation of the quasiparticle and optical properties of materials and
  nanostructures}. \emph{Comput. Phys. Commun.} \textbf{2012}, \emph{183},
  1269--1289\relax
\mciteBstWouldAddEndPuncttrue
\mciteSetBstMidEndSepPunct{\mcitedefaultmidpunct}
{\mcitedefaultendpunct}{\mcitedefaultseppunct}\relax
\EndOfBibitem
\bibitem[Ismail-Beigi(2006)]{IsmailBeigi2006}
Ismail-Beigi,~S. Truncation of periodic image interactions for confined
  systems. \emph{Phys. Rev. B} \textbf{2006}, \emph{73}, 233103\relax
\mciteBstWouldAddEndPuncttrue
\mciteSetBstMidEndSepPunct{\mcitedefaultmidpunct}
{\mcitedefaultendpunct}{\mcitedefaultseppunct}\relax
\EndOfBibitem
\bibitem[Deslippe \latin{et~al.}(2013)Deslippe, Samsonidze, Jain, Cohen, and
  Louie]{Deslippe2013}
Deslippe,~J.; Samsonidze,~G.; Jain,~M.; Cohen,~M.~L.; Louie,~S.~G. Coulomb-hole
  summations and energies for $GW$ calculations with limited number of empty
  orbitals: A modified static remainder approach. \emph{Phys. Rev. B}
  \textbf{2013}, \emph{87}, 165124\relax
\mciteBstWouldAddEndPuncttrue
\mciteSetBstMidEndSepPunct{\mcitedefaultmidpunct}
{\mcitedefaultendpunct}{\mcitedefaultseppunct}\relax
\EndOfBibitem
\bibitem[Golze \latin{et~al.}(2019)Golze, Dvorak, and Rinke]{GDR19}
Golze,~D.; Dvorak,~M.; Rinke,~P. The $GW$ Compendium: A Practical Guide to
  Theoretical Photoemission Spectroscopy. \emph{Front. Chem.} \textbf{2019},
  \emph{7}, 377\relax
\mciteBstWouldAddEndPuncttrue
\mciteSetBstMidEndSepPunct{\mcitedefaultmidpunct}
{\mcitedefaultendpunct}{\mcitedefaultseppunct}\relax
\EndOfBibitem
\bibitem[{\"O}z{\c c}elik \latin{et~al.}(2020){\"O}z{\c c}elik, Li, Xiong, and
  Paesani]{Oezcelik2020}
{\"O}z{\c c}elik,~V.~O.; Li,~Y.; Xiong,~W.; Paesani,~F. Modeling Spontaneous
  Charge Transfer at Metal/Organic Hybrid Heterostructures. \emph{J. Phys.
  Chem. C} \textbf{2020}, \emph{124}, 4802--4809\relax
\mciteBstWouldAddEndPuncttrue
\mciteSetBstMidEndSepPunct{\mcitedefaultmidpunct}
{\mcitedefaultendpunct}{\mcitedefaultseppunct}\relax
\EndOfBibitem
\bibitem[Sato \latin{et~al.}(1981)Sato, Seki, and Inokuchi]{Sato1981}
Sato,~N.; Seki,~K.; Inokuchi,~H. Polarization energies of organic solids
  determined by ultraviolet photoelectron spectroscopy. \emph{J. Chem. Soc.}
  \textbf{1981}, \emph{77}, 1621--1633\relax
\mciteBstWouldAddEndPuncttrue
\mciteSetBstMidEndSepPunct{\mcitedefaultmidpunct}
{\mcitedefaultendpunct}{\mcitedefaultseppunct}\relax
\EndOfBibitem
\bibitem[Refaely-Abramson \latin{et~al.}(2013)Refaely-Abramson, Sharifzadeh,
  Jain, Baer, Neaton, and Kronik]{RefaelyAbramson2013}
Refaely-Abramson,~S.; Sharifzadeh,~S.; Jain,~M.; Baer,~R.; Neaton,~J.~B.;
  Kronik,~L. Gap renormalization of molecular crystals from density-functional
  theory. \emph{Phys. Rev. B} \textbf{2013}, \emph{88}, 081204\relax
\mciteBstWouldAddEndPuncttrue
\mciteSetBstMidEndSepPunct{\mcitedefaultmidpunct}
{\mcitedefaultendpunct}{\mcitedefaultseppunct}\relax
\EndOfBibitem
\bibitem[Baran \latin{et~al.}(2017)Baran, Ashraf, Hanifi, Abdelsamie,
  Gasparini, R{\"o}hr, Holliday, Wadsworth, Lockett, Neophytou, Emmott, Nelson,
  Brabec, Amassian, Salleo, Kirchartz, Durrant, and McCulloch]{Baran2017}
Baran,~D.; Ashraf,~R.~S.; Hanifi,~D.~A.; Abdelsamie,~M.; Gasparini,~N.;
  R{\"o}hr,~J.~A.; Holliday,~S.; Wadsworth,~A.; Lockett,~S.; Neophytou,~M.;
  Emmott,~C. J.~M.; Nelson,~J.; Brabec,~C.~J.; Amassian,~A.; Salleo,~A.;
  Kirchartz,~T.; Durrant,~J.~R.; McCulloch,~I. Reducing the
  efficiency-stability-cost gap of organic photovoltaics with highly efficient
  and stable small molecule acceptor ternary solar cells. \emph{Nat. Mater.}
  \textbf{2017}, \emph{16}, 363--369\relax
\mciteBstWouldAddEndPuncttrue
\mciteSetBstMidEndSepPunct{\mcitedefaultmidpunct}
{\mcitedefaultendpunct}{\mcitedefaultseppunct}\relax
\EndOfBibitem
\bibitem[Kroon \latin{et~al.}(2008)Kroon, Lenes, Hummelen, Blom, and
  de~Boer]{Kroon2008}
Kroon,~R.; Lenes,~M.; Hummelen,~J.~C.; Blom,~P. W.~M.; de~Boer,~B. Small
  Bandgap Polymers for Organic Solar Cells (Polymer Material Development in the
  Last 5 Years). \emph{Polymer Rev.} \textbf{2008}, \emph{48}, 531--582\relax
\mciteBstWouldAddEndPuncttrue
\mciteSetBstMidEndSepPunct{\mcitedefaultmidpunct}
{\mcitedefaultendpunct}{\mcitedefaultseppunct}\relax
\EndOfBibitem
\bibitem[Hou \latin{et~al.}(2006)Hou, Tan, Yan, He, Yang, and Li]{Hou2006}
Hou,~J.; Tan,~Z.; Yan,~Y.; He,~Y.; Yang,~C.; Li,~Y. Synthesis and Photovoltaic
  Properties of Two-Dimensional Conjugated Polythiophenes with
  Bi(thienylenevinylene) Side Chains. \emph{J. Am. Chem. Soc.} \textbf{2006},
  \emph{128}, 4911--4916\relax
\mciteBstWouldAddEndPuncttrue
\mciteSetBstMidEndSepPunct{\mcitedefaultmidpunct}
{\mcitedefaultendpunct}{\mcitedefaultseppunct}\relax
\EndOfBibitem
\bibitem[Qian \latin{et~al.}(2015)Qian, Umari, and Marzari]{Qian2015}
Qian,~X.; Umari,~P.; Marzari,~N. First-principles investigation of organic
  photovoltaic materials ${\text{C}}_{60}, {\text{C}}_{70},
  [{\text{C}}_{60}]\text{PCBM}$, and bis-$[{\text{C}}_{60}]\text{PCBM}$ using a
  many-body ${G}_{0}{W}_{0}$-Lanczos approach. \emph{Phys. Rev. B}
  \textbf{2015}, \emph{91}, 245105\relax
\mciteBstWouldAddEndPuncttrue
\mciteSetBstMidEndSepPunct{\mcitedefaultmidpunct}
{\mcitedefaultendpunct}{\mcitedefaultseppunct}\relax
\EndOfBibitem
\bibitem[Li(2012)]{Li2012}
Li,~Y. Molecular Design of Photovoltaic Materials for Polymer Solar Cells:
  Toward Suitable Electronic Energy Levels and Broad Absorption. \emph{Acc.
  Chem. Res.} \textbf{2012}, \emph{45}, 723--733\relax
\mciteBstWouldAddEndPuncttrue
\mciteSetBstMidEndSepPunct{\mcitedefaultmidpunct}
{\mcitedefaultendpunct}{\mcitedefaultseppunct}\relax
\EndOfBibitem
\bibitem[Shih \latin{et~al.}(2013)Shih, Huang, Lin, Li, Chen, Chiu, and
  Chen]{Shih2013}
Shih,~M.-C.; Huang,~B.-C.; Lin,~C.-C.; Li,~S.-S.; Chen,~H.-A.; Chiu,~Y.-P.;
  Chen,~C.-W. Atomic-Scale Interfacial Band Mapping across Vertically
  Phased-Separated Polymer/Fullerene Hybrid Solar Cells. \emph{Nano Lett.}
  \textbf{2013}, \emph{13}, 2387--2392\relax
\mciteBstWouldAddEndPuncttrue
\mciteSetBstMidEndSepPunct{\mcitedefaultmidpunct}
{\mcitedefaultendpunct}{\mcitedefaultseppunct}\relax
\EndOfBibitem
\bibitem[Ihn(2010)]{2010}
Ihn,~T. \emph{Semiconductor Nanostructures: Quantum states and electronic
  transport}; 1st ed. Oxford University Press, 2010\relax
\mciteBstWouldAddEndPuncttrue
\mciteSetBstMidEndSepPunct{\mcitedefaultmidpunct}
{\mcitedefaultendpunct}{\mcitedefaultseppunct}\relax
\EndOfBibitem
\bibitem[Lo \latin{et~al.}(2011)Lo, Mirkovic, Chuang, Burda, and
  Scholes]{Lo2011}
Lo,~S.~S.; Mirkovic,~T.; Chuang,~C.-H.; Burda,~C.; Scholes,~G.~D. Emergent
  Properties Resulting from Type-II Band Alignment in Semiconductor
  Nanoheterostructures. \emph{Adv. Mater.} \textbf{2011}, \emph{23},
  180--197\relax
\mciteBstWouldAddEndPuncttrue
\mciteSetBstMidEndSepPunct{\mcitedefaultmidpunct}
{\mcitedefaultendpunct}{\mcitedefaultseppunct}\relax
\EndOfBibitem
\bibitem[Ameri \latin{et~al.}(2013)Ameri, Khoram, Min, and Brabec]{Ameri2013}
Ameri,~T.; Khoram,~P.; Min,~J.; Brabec,~C.~J. Organic Ternary Solar Cells: A
  Review. \emph{Adv. Mater.} \textbf{2013}, \emph{25}, 4245--4266\relax
\mciteBstWouldAddEndPuncttrue
\mciteSetBstMidEndSepPunct{\mcitedefaultmidpunct}
{\mcitedefaultendpunct}{\mcitedefaultseppunct}\relax
\EndOfBibitem
\bibitem[Br\'{e}das \latin{et~al.}(2004)Br\'{e}das, Beljonne, Coropceanu, and
  Cornil]{BBCC04}
Br\'{e}das,~J.-L.; Beljonne,~D.; Coropceanu,~V.; Cornil,~J. Charge-Transfer and
  Energy-Transfer Processes in $\pi$-Conjugated Oligomers and Polymers: A
  Molecular Picture. \emph{Chem. Rev.} \textbf{2004}, \emph{104},
  4971--5003\relax
\mciteBstWouldAddEndPuncttrue
\mciteSetBstMidEndSepPunct{\mcitedefaultmidpunct}
{\mcitedefaultendpunct}{\mcitedefaultseppunct}\relax
\EndOfBibitem
\bibitem[Tamblyn \latin{et~al.}(2011)Tamblyn, Darancet, Quek, Bonev, and
  Neaton]{Tamblyn2011}
Tamblyn,~I.; Darancet,~P.; Quek,~S.~Y.; Bonev,~S.~A.; Neaton,~J.~B. Electronic
  energy level alignment at metal-molecule interfaces with a $GW$ approach.
  \emph{Phys. Rev. B} \textbf{2011}, \emph{84}, 201402\relax
\mciteBstWouldAddEndPuncttrue
\mciteSetBstMidEndSepPunct{\mcitedefaultmidpunct}
{\mcitedefaultendpunct}{\mcitedefaultseppunct}\relax
\EndOfBibitem
\bibitem[Chen \latin{et~al.}(2017)Chen, Tamblyn, and Quek]{Chen2017}
Chen,~Y.; Tamblyn,~I.; Quek,~S.~Y. Energy Level Alignment at Hybridized
  Organic--Metal Interfaces: The Role of Many-Electron Effects. \emph{J. Phys.
  Chem. C} \textbf{2017}, \emph{121}, 13125--13134\relax
\mciteBstWouldAddEndPuncttrue
\mciteSetBstMidEndSepPunct{\mcitedefaultmidpunct}
{\mcitedefaultendpunct}{\mcitedefaultseppunct}\relax
\EndOfBibitem
\bibitem[Liu(2020)]{Liu2020}
Liu,~Z.-F. Dielectric embedding GW for weakly coupled molecule-metal
  interfaces. \emph{J. Chem. Phys.} \textbf{2020}, \emph{152}, 054103\relax
\mciteBstWouldAddEndPuncttrue
\mciteSetBstMidEndSepPunct{\mcitedefaultmidpunct}
{\mcitedefaultendpunct}{\mcitedefaultseppunct}\relax
\EndOfBibitem
\end{mcitethebibliography}
\end{document}